\documentclass[aps,prfluids,twocolumn,longbibliography,groupedaddress]{revtex4-1}
\usepackage[dvipdfmx]{graphicx}
\usepackage[dvipdfmx]{color}
\usepackage{amssymb}
\usepackage{amsmath}
\usepackage{amsfonts}

\begin{document}

\title{Obstacle-shape effect in a two-dimensional granular silo flow field}
\author{K. Endo$^1$, K. Anki Reddy$^2$, and H. Katsuragi$^1$}
\affiliation{$^1$Department of Earth and Environmental Sciences, Nagoya University, Nagoya 464-8601, Japan \\
$^2$Department of Chemical Engineering, Indian Institute of Technology Guwahati, Guwahati 781039, India}

\date{\today}

\begin{abstract}
We conducted simple experiment and numerical simulation of two-dimensional granular discharge flow driven by gravity under the influence of an obstacle. According to the previous work (Zuriguel {\it et al.,\,Phys.\,Rev.\,Lett.}\,{\bf 107}: 278001, 2011), the clogging of granular discharge flow can be suppressed by putting a circular obstacle at a proper position. In order to investigate the details of obstacle effect in granular flow, we focused on particle dynamics in this study. From the experimental and numerical data, we found that the obstacle remarkably affects the horizontal-velocity distribution and packing fraction at the vicinity of the exit. In addition to the circular obstacle, we utilized triangular, inverted-triangular, and horizontal-bar obstacles to discuss the obstacle-shape effect in granular discharge flow. Based on the investigation of dynamical quantities such as velocity distributions, granular temperature, and volume fraction, we found that the triangular obstacle or horizontal bar could be very effective to prevent the clogging. From the obtained result, we consider that the detouring of particles around the obstacle and resultant low packing fraction at the exit region effectively prevent the clogging in a certain class of granular discharge flow.
\end{abstract}


\maketitle

\section{Introduction}
\label{sec:intro}
Gravity-driven granular silo flow shows some intriguing phenomena. Its behavior is quite different from usual fluid behavior. For instance, discharge flow rate of gravity-driven granular silo flow is independent of its height (thickness) in the silo. Such a constant flow cannot be achieved when usual fluid is used. In usual fluid, the discharge flow rate varies depending on the layer's height~(Torricelli's theorem~\cite{Faber1995}). The pressure structure within a granular silo tends to saturate in deep part due to Janssen effect~\cite{Janssen1985}. This saturated (constant) pressure could be a reason for the constant flow rate. However, according to the recent study~\cite{Aguirre:2010ho}, the constant granular flow might not relate to Janssen effect. Namely, the origin of steadiness in granular silo flow has not yet been fully understood. Beverloo {\it et al.} experimentally obtained the scaling for the flow rate in granular silo flow which depends on the sizes of particles and exit~\cite{Beverloo1961}. In particular, when the exit width $W$ decreases to a certain limit (e.g. $W\simeq6D$, where $D$ is the particle diameter), an arch structure of the granular particles could easily be formed around the exit~\cite{Helbing2006}. Due to this arch formation, particles suddenly clog to arrest the flow~\cite{To2001,To2005,Zuriguel2005,Janda2008}. The clogging phenomenon is often a severe problem in transporting granular materials in industry. 

An obstacle placed within the granular silo flow could affect the flow property such as clogging condition. In fact, the clogging probability is decreased by inserting an obstacle into the granular silo at a proper position~\cite{Zuriguel2011,Lozano2012,Zuriguel2014sr}. The clogging phenomena have also been observed in crowd and animal flows, and obstacles have been used to control these flows as well~\cite{Helbing2005,Frank2011,Garcimartin2015}. By the experimental simulation of evacuating exit flow, it has been revealed that the obstacle set in front of the exit can decrease the outflow period~\cite{Kawaguchi2012,Zuriguel2014sr}. Although the vibration or concentrated air flow can also be utilized to avoid the granular clogging~\cite{Mankoc2009,Zuriguel2005}, obstacle usage is much easier than vibration and air flow in various situations. 

Arch formation by particles at the exit region is the fundamental process to proceed to the clogging. The smaller the exit size, the larger the clogging probability becomes. Recently, non-zero clogging probability even for very large exit has been proposed on the basis of experimental result~\cite{Thomas:2015hv}. The self-similar density and velocity profiles around the exit have also been observed~\cite{Janda2012}. In addition, velocity field in granular silo flow without an obstacle has been studied in various works~\cite{Choi2004,Moka:2005jx,Orpe2007,Thomas:2016bt}. However, the characteristics of granular-flow field under the influence of obstacle has not yet been understood well. Moreover, details of flow-field conditions must be influenced by the shape of obstacle and/or particles as well. Although the particle-shape dependence of granular silo flow has recently been studied~\cite{Ashour:2017hb}, the effect of shape of obstacle has not yet been studied well so far. 

Therefore, we conducted simple experiment and numerical simulation of two-dimensional granular silo flow driven by gravity under the influence of an obstacle. In the experiment, we traced individual particles from the flow images acquired by a high-speed camera. Numerical simulation of granular silo flow has also been performed to confirm the major effect of obstacle to prevent the clogging. Particularly, we mainly compare the various obstacle-shape results in order to discuss the physical origin of clogging reduction by obstacle.

\section{Methods}
\subsection{Experiment}
\label{sec:experiment}
\begin{figure}
\begin{center}
  \includegraphics[width=80mm]{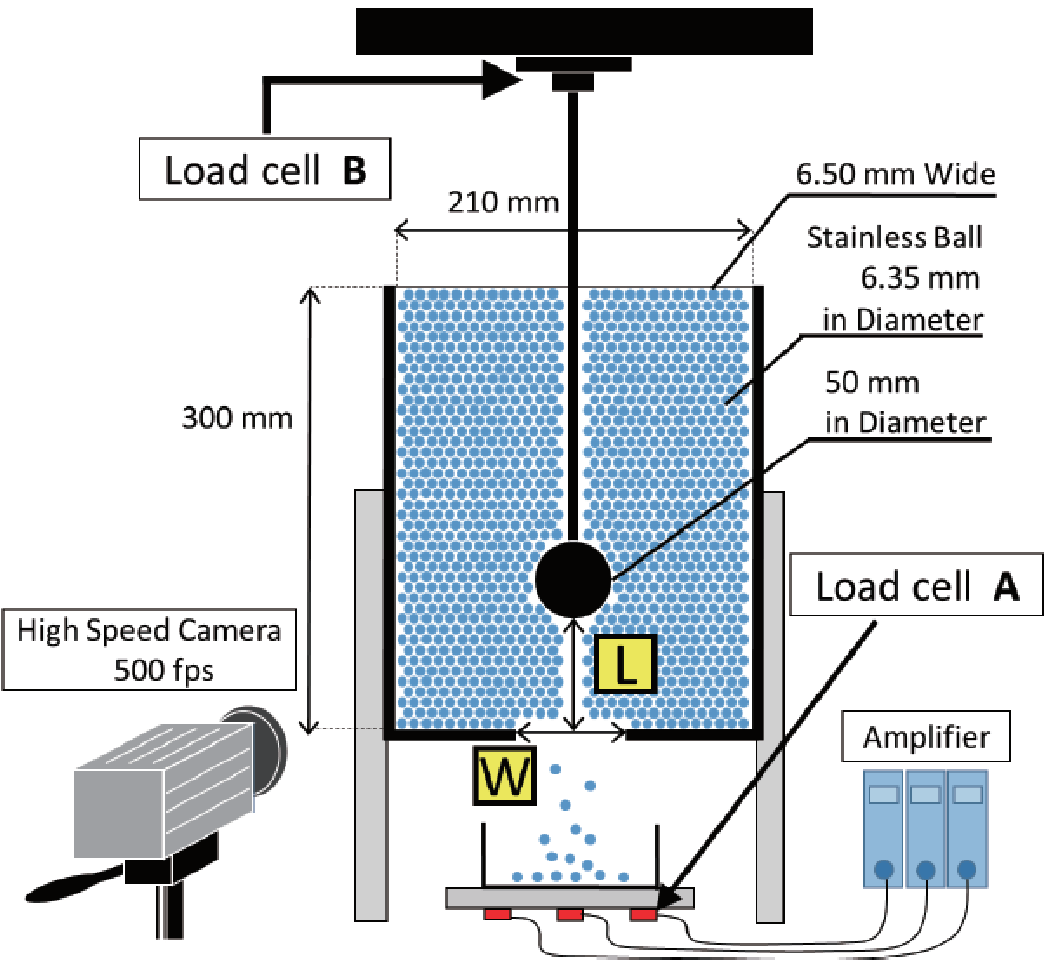} 
  \includegraphics[width=80mm]{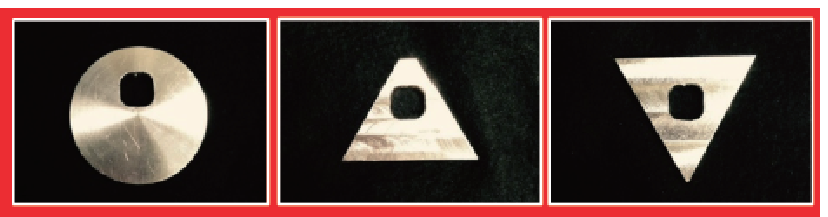}
\end{center}
\caption{Experimental apparatus and obstacles. A schematic diagram of the experimental setup is shown above. Width of the exit $W$ and distance between the exit and obstacle $L$ are principal parameters in this system. Below images show obstacles used in this experiment: a circle of 50~mm in diameter, an equilateral triangle of one side 50~mm, and an inverted equilateral triangle of one side 50~mm. Holes in the obstacles were opened for making attaching mechanism.}
\label{fig:apparatus}       
\end{figure}

An experimental system we built and pictures of obstacles used are shown in Fig.~\ref{fig:apparatus}.
In this experiment, we prepare a two-dimensional cell consisting of two acrylic transparent plates and aluminum rectangles for the side and bottom walls.
The inner dimension of the cell is 6.50 $\times$ 210 $\times$ 300~mm (thickness $\times$ width $\times$ height). 
The obstacles are made of stainless steel and have 6.0~mm thickness.
We use three types of obstacles : a circle of 50~mm diameter, an equilateral triangle of one side 50~mm, and an inverted equilateral triangle of the identical size. 
After inserting an obstacle and filling the cell with stainless-steel particles having 6.35 $\pm$ 0.05~mm diameter, a discharge flow is triggered by opening a small exit at the center of the cell's bottom. We do not refill the container by particles. Namely, the measurement lasts until all the particles are discharged or the clogging occurrence. 

Discharged particles are caught by a container placed below the cell.
Load cell sensors (LMB-A, KYOWA) are used to measure the mass of discharged particles.
The obstacle is fixed to a stainless-steel pole of 6.0~mm diameter which is connected to an universal testing machine (AG-X, SHIMADZU) to control the position of the obstacle. Drag force exerting on the obstacle can also be measured using the testing machine. However, here we focus on the clogging problem in this paper. The drag force characterization will be presented elsewhere~\cite{Katsuragi2016}.
Flow rate data are taken at 100 Samples/s sampling rate.
We also measure the microscopic granular flow field by using a high-speed camera (FASTCAM SA5, Photron).
The acquired images are analyzed by means of particle tracking velocimetry (PTV) method implemented by LabVIEW software (vision toolkit). 
The frame rate is fixed at 500~fps.
The spatial resolution of the images is 0.26~mm/pixel, and the images consist of $832 \times 880$ pixels corresponding to the field of view $220 \times 233$~mm$^2$. 

The main parameters in this experiment are the width of the exit $W$ and the vertical distance between the exit and obstacle $L$.
Specifically, $W$ is varied as 25, 30, 40 and 60~mm. $L$ is changed by 5~mm from $L=0$~mm to $L=50$~mm and by 10~mm from $L=50$~mm to $L=100$~mm.
The case without any obstacle is denoted by $L=\infty$.

\subsection{Simulation methodology}
\label{sec:numerical}

We used the contact force models based on~\cite{Brilliantov1996} and explained further in~\cite{Silbert2001} to simulate the gravity driven flow of granular media. Suppose if particle \textit{i} is in contact with particle \textit{j}, the contact force on particle \textit{i} resolved into two components (along the line joining centers and perpendicular to the line joining centers) can be expressed as $\textbf{F}_{ij} = \textbf{F}_{ij}^{n} + \textbf{F}_{ij}^{t}$. For the present model,  the normal and tangential force components are given by $\textbf{F}_{ij}^{n} = \sqrt{\delta_{ij}} \sqrt{\frac{R_{i}R_{j}}{R_{i}+R_{j}}}(k_{n}\delta_{ij}\textbf{n}_{ij} - m_{\textup{eff}}\gamma_{n}\textbf{v}_{ij}^{n}) $ and $\textbf{F}_{ij}^{t} = \sqrt{\delta_{ij}}\sqrt{\frac{R_{i}R_{j}}{R_{i}+R_{j}}}(k_{t}\textbf{u}_{ij}^{t} - m_{\textup{eff}} \gamma_{t} \textbf{v}_{ij}^{t})$, respectively. Here $R_{i}$ and $R_{j}$ are the radii of particles $i$ and $j$.   If particle \textit{i} and particle \textit{j} respectively have positions $\textbf{r}_{i}$ and $\textbf{r}_{j}$, masses $m_{i}$ and $m_{j}$, linear velocities $\textbf{v}_{i}$ and $\textbf{v}_{j}$,
 angular velocities $\textbf{w}_{i}$ and $\textbf{w}_{j}$,
 quantities used in the above force equation are defined as $\delta_{ij}=d-|\textbf{r}_{i}-\textbf{r}_{j}|$, $\textbf{v}_{ij}^{n}=(\textbf{v}_{ij}\cdot\textbf{n}_{ij})\textbf{n}_{ij}$, $\textbf{v}_{ij}=\textbf{v}_{i}-\textbf{v}_{j}$, 
$\textbf{v}_{ij}^{t}=\textbf{v}_{ij}-\textbf{v}_{ij}^{n}-\frac{1}{2}\textbf{r}_{ij} \times (\textbf{w}_{i}+\textbf{w}_{j})$, $\textbf{n}_{ij}=\textbf{r}_{ij}/|\textbf{r}_{ij}|$ and $m_{\textup{eff}}=m_{i}m_{j}/(m_{i}+m_{j})$. 

Once the force acting on particle \textit{i} is known, classical equations of motion can be integrated with the suitable numerical technique to update its position and velocities.  In the simulation, elastic constant for normal contact  ($k_{n}$) is $2\times10^{8} mg/d^{2}$, whereas the the elastic constant in tangential direction ($k_{t}$) is $\frac{2}{7}k_{n}$~\cite{Silbert2001}. Here, $m$ is the mass and $d$ is the diameter of the particle and $g$ is the acceleration due to gravity. The viscoelastic  damping constant for normal contact  ($\gamma_{n}$) is $1850 \sqrt{g}/d^{1.5}$ and for tangential contact $\gamma_{t}$ is half of the $\gamma_{n}$. The chosen constants represent the properties of material used in the experimental study. It is important to mention that we included memory effect in the modeling of tangential forces. Tangential displacement  between  the two contact particles $(\textbf{u}_{ij}^{t})$ will be computed from the initiation of the contact (it will be zero at the initiation of contact) and it is adjusted so that the local yield criterion $|\textbf{F}_{ij}^{t}| < |\mu \textbf{F}_{ij}^{n}|$ is satisfied. Rate of change of this elastic tangential displacement  $(\textbf{u}_{ij}^{t})$ is given by ${d\textbf{u}_{ij}^{t}}/{dt} = \textbf{v}_{ij}^{t} - (\textbf{u}_{ij}^{t}\cdot\textbf{v}_{ij})\textbf{r}_{ij}/|\textbf{r}_{ij}|^{2}$. In the present simulations, the friction coefficient $\mu$  has been set equal to 0.36 for particle-particle interactions and 0.5 for particle-wall interactions. The timestep in the present simulation is  0.000002. We arrived at this value of time step by computing the collision time with the help of equations given in~\cite{Schwager1998,Brilliantov1996}  and this collision time was found to be  $O(10^{-4})$. All the lengths are scaled by particle diameter ($d$), which is $1.0$ in the present case. Time, velocities and forces are measured in units of $\sqrt{d/g}$, $\sqrt{g d}$, and $mg$, respectively. Elastic constants and viscoelastic damping constants are measured in units of $(mg/d^{2})$ and $\frac{1}{(\mbox{time}*\mbox{distance})}$. Density of the particles is also set as $1.0$.  Obstacle shape in the present simulation is made using  spherical particles of diameter 1.0. For example, to create a circular obstacle of diameter $8 d$, we need approximately 22 particles to make the outer layer of circle. Similarly, to make a triangular obstacle which has length of side as $8 d$, we need 36 particles.

\begin{figure}
\begin{center}
\includegraphics[width=80mm]{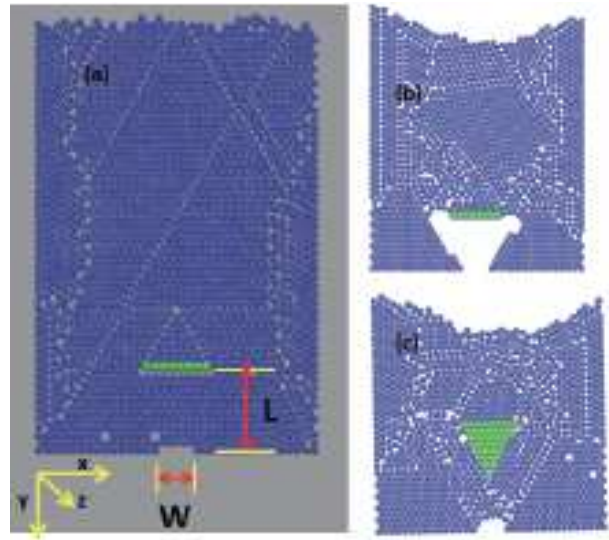}
\caption{(a)~Simulation system with a horizontal-bar obstacle consisting of green-color spheres. Typical clogging configurations observed in simulations for (b)~the flow past horizontal-bar obstacle, $L=45$ and $W=25$~mm and for (c)~the flow past inverted-triangle obstacle, $L=60$ and $W=25$~mm.}
\label{fig:simuexample}
\end{center}
\end{figure}

Initial configuration is prepared by pouring the particles into a container under the gravity. There are 1900 particles in the simulation which is in correspondence with the number of particles used in experimental study discussed in this paper. This dimension of a simulation box is $34d \times 48d \times 1d$.  The simulation system is as shown in Fig.~\ref{fig:simuexample}. Flat frictional walls are present in the $x$-direction and periodic boundary conditions are applied in the $z$-direction. Bottom wall in the $y$-direction is made up of spherical particles of diameter 1.0 $d$ and to simulate the discharge of particles through orifice, certain number of spheres will be removed from the  bottom $y$-wall (For $W=25$~mm, we need to remove 4 spheres from the bottom wall). All the simulations are carried out using the Large Atomic Molecular Massively Parallel Simulator (LAMMPS)~\cite{Plimpton1995}. In the experiment, the mean grain diameter is 6.35 mm. To make a comparison between the numerical and experimental data, $L$ and $W$ values in the simulation are transformed to the real length scale by multiplying the grain diameter with 6.35 mm in the following data plots. Furthermore, the elastic constant for normal contact used in the simulation results in Young's modulus $(Y)$ of $108.5$~GPa for a stainless steel spherical particle of diameter 6.35 mm if we use the equation $k_{n} = \frac{2Y}{3(1-\nu^{2})}$. Here density and Poisson's ratio ($\nu$) for stainless steel are taken as $8000$~kg/m$^{3}$ and $0.275$. These values are more or less reasonable for the steel.

\section{Experimental results and analyses}
\label{sec:expresults}
\subsection{Clogging diagram}
\label{subsec:Phase_diagram}
\begin{figure}
\begin{center}
  \includegraphics[width=80mm]{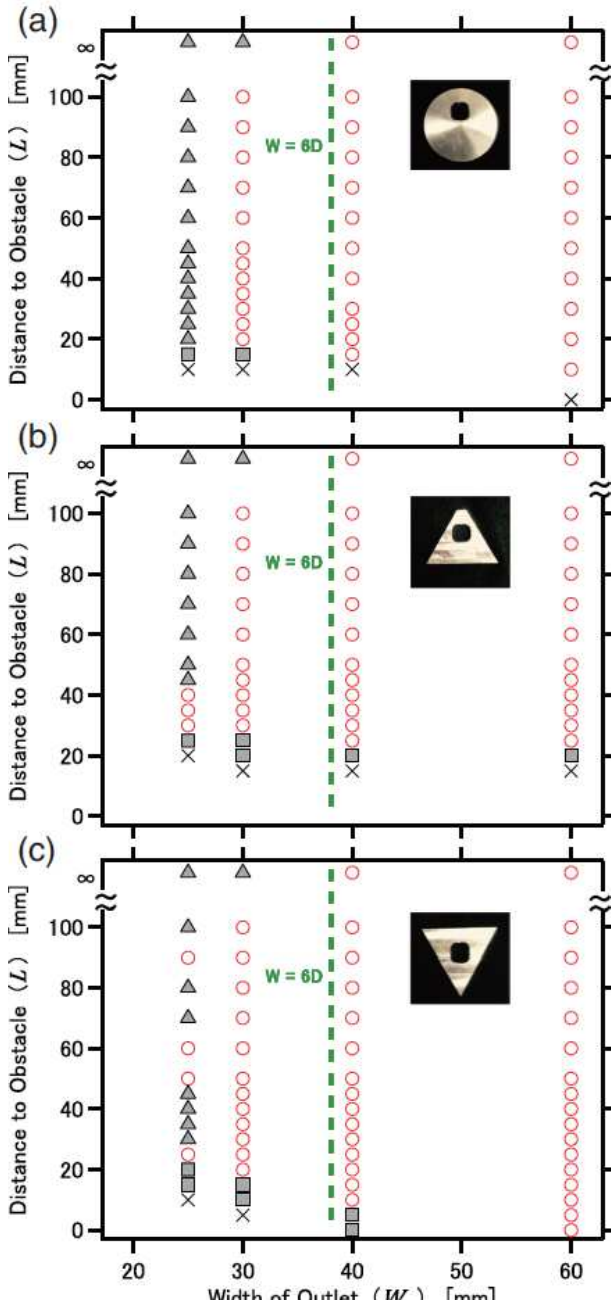} 
\end{center}
\caption{Clogging diagrams for (a) circular, (b) triangular, and (c) inverted-triangular obstacle cases.
Experiments were conducted three or more times (in $W\leq$ 40~mm) to make this diagram.
$L={\infty}$ means the case without an obstacle.
Circular symbols indicate that the clogging did not occur.  Triangular symbols indicate that the clogging by arching at the exit~(Fig.~\ref{fig:clog}(a)) occurred. Square symbols indicate that the clogging occurred between the obstacle and the bottom wall~(Fig.~\ref{fig:clog}(b)), and $\times$ indicates that granular flow did not occur at all.
Green broken line corresponds to $W=6D$. }
\label{fig:phasediagram} 
\end{figure}

First, the clogging-occurrence conditions depending on $W$ and $L$ are examined. In Fig.~\ref{fig:phasediagram}, obtained clogging diagrams for (a) circular, (b) triangular, and (c) inverted-triangular cases are shown. Three or more experimental runs are performed for each experimental condition. If the flow clogs during discharge at least once, that condition is regarded as a clogging case. On the other hand, if all the particles are discharged without clogging for all experimental runs, that condition is assigned to the no-clogging case. In Fig.~\ref{fig:phasediagram}, no clogging is expressed by circular marks while triangle, square, and cross symbols indicate the clogging conditions. 
In this study, the number of experimental runs might not be sufficient to precisely discuss the clogging probability. Moreover, we did not compute the avalanche size distribution. Thus, the clogging diagrams shown in Fig.~\ref{fig:phasediagram} are more or less qualitative ones. Therefore, the quantitative analysis for the clogging probability would not be discussed in this study. However, some characteristic features can be confirmed in these diagrams as discussed below.

\begin{figure}
\begin{center}
  \includegraphics[width=80mm]{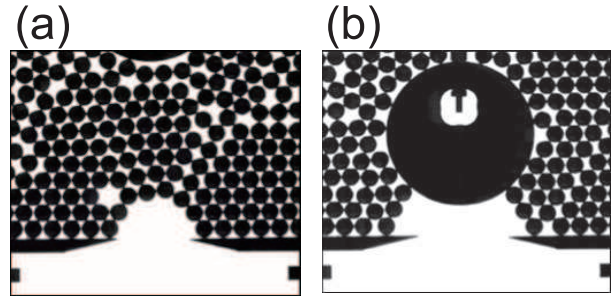} 
\end{center}
\caption{(a) Clogging by arch formation around the exit. (b)~Clogging by arch formation between the obstacle and bottom wall.}
\label{fig:clog}       
\end{figure}

As shown in Figs.~\ref{fig:simuexample}(c) and \ref{fig:clog}(a), the clogging is basically induced by an arch structure formed at the exit region. If an obstacle is too close to the exit (i.e., $L$ is too small), arch formation occurs between the obstacle and bottom wall~(Fig.~\ref{fig:clog}(b)). This type of arch formation is different from the ordinary one which is formed around the exit~(Fig.\ref{fig:clog}(a)). In the clogging diagram~(Fig.~\ref{fig:phasediagram}), the arch formation around the exit is denoted by triangle symbols, and the arch formation between the obstacle and bottom wall is presented by square symbols. If the flow is not induced from the very initial state, cross symbols are assigned. In the numerical simulation, a slightly different clogging mode (Fig.~\ref{fig:simuexample}(b)) is also observed. In this clogging mode, the arch is formed around the obstacle edge and the stable triangular lattice slope. Although this could be regarded as a type of obstacle-wall clogging, this mode cannot be observed in the experiment since it is difficult to make the stable triangular lattice slope in the actual experiment.

When $W$ is large enough, the flow is relatively smooth and difficult to clog. By contrast, when $W=25$~mm, clogging frequently occurs. In Fig.~\ref{fig:phasediagram}(a) (circle case), $L$ dependence of the clogging occurence is not clear. In Fig.~\ref{fig:phasediagram}(c) (inverted-triangle case), although some no-clogging states can be found in $W=25$~mm, any clear trend of $L$-dependent clogging occurence cannot be confirmed. Contrastively, in Fig.~\ref{fig:phasediagram}(b) (triangle case), clear reduction of clogging occurence can be confirmed in the relatively small $L$ regime ($30$~mm $ \leq L \leq 40$~mm). This anti-clogging tendency due to the obstacle is qualitatively consistent with the previous work using a circular obstacle~\cite{Zuriguel2011}. In this study, the anti-clogging by putting an obstacle can clearly be confirmed particularly in the triangle case.

From the above-mentioned observations, we consider that the flow situation could significantly depend on the shape of the obstacle. At $W=30$~mm, clogging due to the arch formation at the exit region occurs only at $L=\infty$ (no-obstacle case). The case of $W=25$~mm is more interesting because the clogging occurence variation depending on $L$ can be observed. The condition to prevent the clogging might be revealed by comparing these cases (no-obstacle, circle, triangle, and inverted triangle) at $W=25$~mm. Therefore, we focus on the case of $W=25$~mm in the following.

\subsection{Granular flow field}
\label{subsec:Flow_field}
In this study, we would like to relate macroscopic behaviors of granular discharge flow to the statistics of motion of individual particles.
Thus, we track particle motions using PTV method.

\begin{figure}
\begin{center}
  \includegraphics[width=70mm]{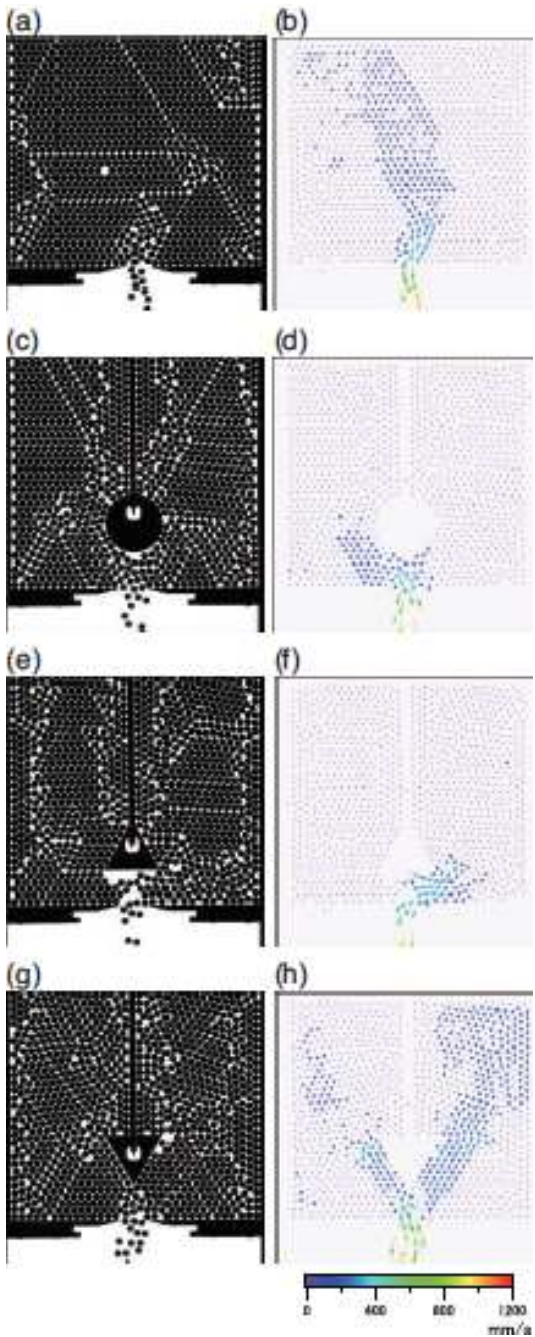} 
\end{center}
\caption{Velocity fields of granular discharge flow acquired by a high-speed camera and computed by PTV method. 
Panel (a) indicates a snapshot of particle configuration without an obstacle. 
Panel (b) is its velocity field computed by each particle's motion in vector representation.
Panel (c) indicates a snapshot of particle configuration with a circular obstacle. 
Panel (d) is its corresponding velocity field．
Panel (e) indicates a snapshot of particle configuration with a triangular obstacle. 
Panel (f) is its corresponding velocity field．
Panel (g) indicates a snapshot of particle configuration with an inverted-triangular obstacle. 
Panel (h) is its corresponding velocity field. Experimental conditions are fixed at $W=25$~mm and $L=30$~mm．
}
\label{fig:flowfield}       
\end{figure}

The examples of velocity fields computed from the series of images of flowing particles acquired by a high-speed camera are shown in Fig.~\ref{fig:flowfield}. Figures~\ref{fig:flowfield}(a,b), (c,d), (e,f), and (g,h) correspond to the cases with no-obstacle, a circle, a triangle, and an inverted triangle, respectively. Experimental conditions in these data are $W=25$~mm and $L=30$~mm, at which the flow is in the marginal state between smooth flow and clogging. Each case has a snapshot of the raw particle image on the left side (Figs.~\ref{fig:flowfield}(a), (c), (e), and (g)) and a corresponding velocity field in vector representation on the right side (Figs.~\ref{fig:flowfield}(b), (d), (f), and (h)).

We first discuss on the no-obstacle case. In Fig.~\ref{fig:flowfield}(b), we can confirm that the vertical component dominates the particle velocities. The relatively large-speed particles distribute at the central zone of the cell (right above the exit). The meandering of flow in this state might relate to the local crystallization due to the monodispersity of particles. Although this velocity field results in the large flow rate, dense flow at the exit region might cause the clogging due to the arch formation. Indeed, the largest flow rate in no-obstacle case has been confirmed~(see the legend in Fig.~\ref{fig:PDF}(a)). It should be noted that all the experiments show steady flow~\cite{Katsuragi2016,Endo2016} even right before the clogging. Although some numerical simulations have reported the increase of the flow rate by the effect of obstacle~\cite{AlonsoMarroquin:2012bd,Murray:2016df}, these studies have simulated the inclined-bottom-wall silo flow. Furthermore, \cite{Murray:2016df} has used the fluctuation of grains and vibration of walls to prevent the clogging. This could significantly affect the flow rate as well. Experiments with horizontal-bottom-wall silo has also shown similar increase of flow rate by the effect of obstacle~\cite{Lozano2012}. However, its increase trend of flow rate was not very significant.

Next, let us focus on the circular-obstacle case. In Fig.~\ref{fig:flowfield}(c,d), the particle configuration and corresponding velocity field of granular flow with a circular obstacle are presented. As can be seen in Fig.~\ref{fig:flowfield}(c), more structural defects in the particles configuration are introduced (compared to no-obstacle case) due to the presence of a large circular obstacle. In addition, the velocity field in this case becomes asymmetric as shown in Fig.~\ref{fig:flowfield}(d). At this moment, particles in the left region are much more active than the other side. Actually, this asymmetry results in the temporal oscillation of the active zone, i.e., the alternate flow is developed. Namely, the obstacle triggers the spatiotemporal inhomogeneity in the granular discharge flow. This inhomogeneity could be a possible reason for avoiding the clogging. However, as shown in the clogging diagram (Fig.~\ref{fig:phasediagram}(a)), the relation between the clogging and the circular obstacle is not so clear. Moreover, we have confirmed that discharge flow rate is almost always steady even in the alternate flow regime~\cite{Endo2016}.

The effect of obstacle is exaggerated by using a triangular obstacle. In Fig.~\ref{fig:flowfield}(e,f), the particle configuration and corresponding velocity field with a triangular obstacle are shown. Qualitative characteristics observed in Fig.~\ref{fig:flowfield}(e,f) are more or less similar to those seen in Fig.~\ref{fig:flowfield}(c,d).
Spatiotemporal inhomogeneity including alternate flow can be induced in this case as well. Moreover, particles-number density (packing fraction) beneath the obstacle (above the exit) significantly decreases in Fig.~\ref{fig:flowfield}(e,f). Since the clogging prevention by the obstacle is most significant in this triangle case, an essential key effect to prevent the clogging must present in this triangular-obstacle case. This point will be discussed later from the viewpoint of velocity distribution and packing fraction.

When we use a triangular obstacle fixed upside down (inverted triangle), different flow field is observed as shown in Fig.~\ref{fig:flowfield}(g,h). Since the angle of triangular shape is commensurate with the triangular lattice structure, particle configuration shows the crystalline structure even in the region between the obstacle and bottom wall. And the converging flow at the exit region can be observed. The flow field is relatively symmetric compared to circle and triangle cases while it still shows a slight asymmetry. Furthermore, the packing fraction right above the exit is not very much reduced in this case.

\subsection{Analyses of particle motions}
\label{subsec:Analyses_of_particle}
\begin{figure}
\begin{center}
  \includegraphics[width=40mm]{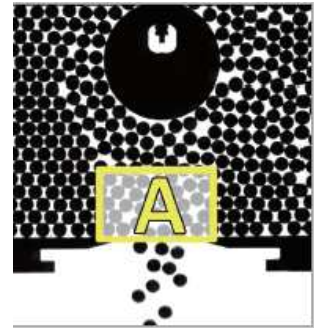} 
\end{center}
\caption{Definition of the focussed region {\sf{A}}.
The yellow square region (region {\sf{A}}) is fixed at the central part above the exit with 50~mm wide and 30~mm high.
When $L$ is smaller than $30$~mm, the height is adjusted to the value of $L$.
}
\label{fig:region}       
\end{figure}

In order to discuss the principal effect for avoiding the clog, here we analyze the region above the exit {\sf{A}} (the yellow square region shown in Fig.~\ref{fig:region}). All data shown in this subsection are based on the analyses in the region {\sf{A}}. The width and height of region {\sf{A}} are basically fixed to 50~mm and 30~mm, respectively. When $L<30$~mm, however, height of the region {\sf{A}} is adjusted to $L$. With respect to the definition of the spatial axis, top left corner in the acquired image corresponds to the origin of vertical ($y$) and horizontal ($x$) axis. And the positive direction of $y$ axis is taken to be the gravitational downward direction.

\subsubsection{Vertical velocity}
\label{subsubsec:Vertical_velocity}

\begin{figure}
\begin{center}
\resizebox{0.45\textwidth}{!}{%
  \includegraphics{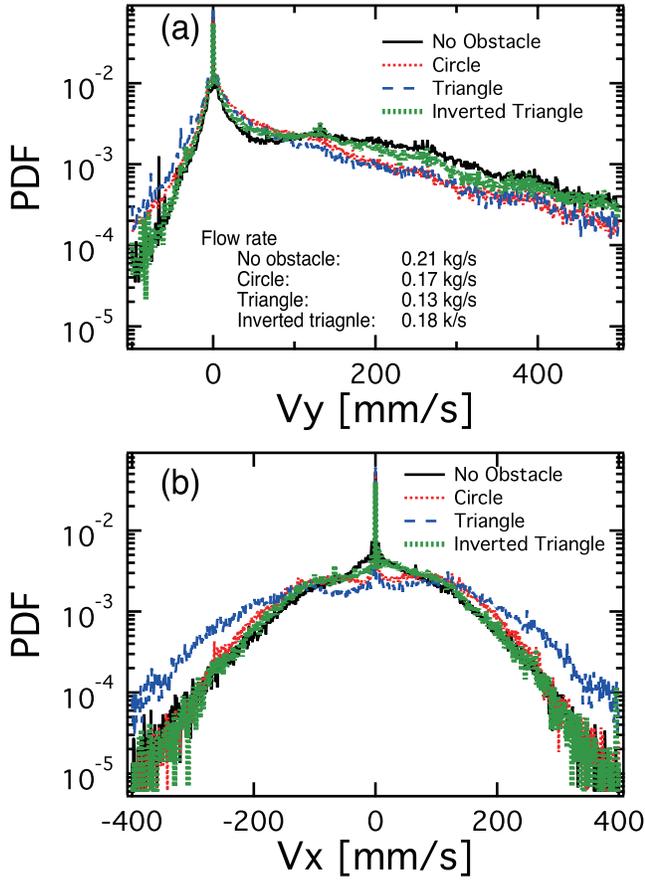} 
}
\end{center}
\caption{
(a) Vertical velocity distribution in the region {\sf{A}}.
PDF of $V_y$ at $W=25$~mm and $L=30$~mm is shown.
The positive direction of $V_y$ corresponds to the gravitational (downward) direction. 
(b) Horizontal velocity distribution in the region {\sf{A}} (PDF of $V_x$ at $W=25$~mm and $L=30$~mm). 
}
\label{fig:PDF}       
\end{figure}

To characterize the velocity field, its probability density function (PDF) is computed from the PTV data. First, the vertical component $V_y$ is focused. Measured PDF of $V_y$ is shown in Fig.~\ref{fig:PDF}(a). In Fig.~\ref{fig:PDF}(a), different colors (and line codes) represent the different obstacle shapes. To reveal the effect of obstacle shape, experimental conditions are fixed at $W=25$~mm and $L=30$~mm. The qualitative form of PDF is basically identical among all the data shown in Fig.~\ref{fig:PDF}(a). In large $V_y$ regime, exponential-like tail can be observed.
This part must principally determine the discharge flow rate. Obviously, the velocity level is reduced by the effect of obstacle. In other words, probability of large $V_y$ in the no-obstacle case is larger than other obstacle cases. The corresponding flow rates are shown in  the legend of Fig.~\ref{fig:PDF}(a). This tendency is consistent with the flow rate reduction due to the obstacle effect~\cite{Katsuragi2016,Endo2016}. Note that, the discharge flow rate is always steady independently of obstacle shapes~\cite{Katsuragi2016,Endo2016}. In Fig.~\ref{fig:PDF}(a), relatively small amount of particles have negative $V_y$. This negative $V_y$ stems from the effective back scattering due to the particle-particle collisions.

\subsubsection{Horizontal velocity and granular temperature}
\label{subsubsec:Horizontal_velocity}
Next, we focus on the behavior of horizontal velocity $V_x$. Figure~\ref{fig:PDF}(b) shows PDF of $V_x$ at $W=25$~mm and $L=30$~mm. These PDFs have the symmetric form because the shape of obstacle is symmetric. The effect of alternate flow is wiped out by the spatiotemporal average to make PDF. The width of PDF depends on the shape of obstacle. In the excited (dilute) granular gas, stretched-exponential-type PDF has been observed in many experiments~\cite{Losert:1999ig,Rouyer:2000ih,vanZon:2004ig}. However, here we are not going to examine the detail structure of PDF because the statistics to discuss the detail shape is limited in this experiment. Instead, we simply discuss the global-shape characterization of the PDF. Specifically, the width of PDF becomes the largest in the triangular-obstacle case while others are somewhat similar in large $|V_x|$ regime. Also, in the small $|V_x|$ regime ($|V_x| \leq 100 $~mm/s), the obstacle dependence in PDF structure can be observed. No-obstacle and inverted-triangular obstacle result in qualitatively similar PDF forms in which the population of smaller $|V_x|$ is always larger than that of larger $|V_x|$. On the other hand, the population dips at relatively small $|V_x|$ regime can be confirmed in the cases of circular and triangular obstacles. Sharp peaks confirmed at $V_x \simeq 0$ come from the almost stopping particles in the silo. Similar peaks can also be found in $V_y$ PDF~(Fig.\ref{fig:PDF}(a)).

\begin{figure}
\begin{center}
\resizebox{0.45\textwidth}{!}{%
  \includegraphics{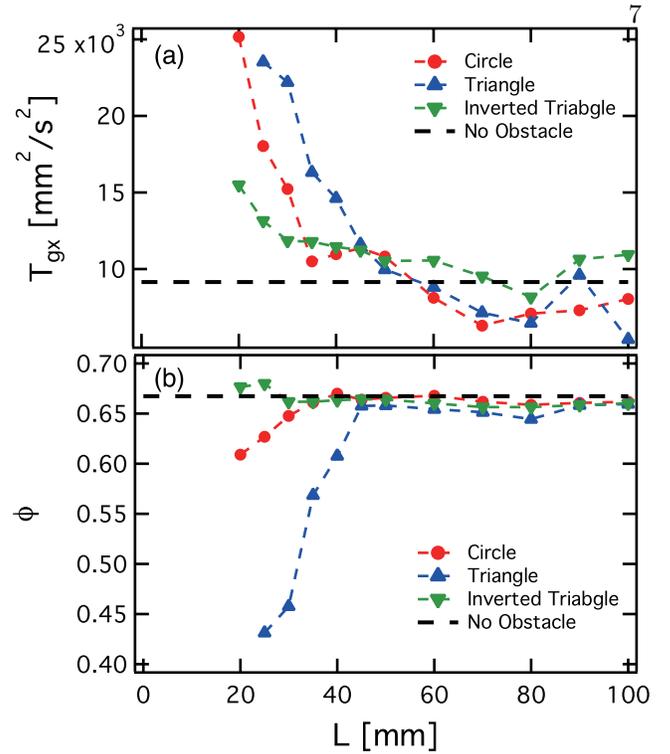} 
}
\end{center}
\caption{
$L$ dependences of (a)~horizontal granular temperature $T_{gx}$ and (b)~packing fraction $\phi$ at $W=25$~mm.
As the obstacle approaches the exit ($L$ becomes small), (a)~$T_{gx}$ increases, and (b)~$\phi$ decreases.
Horizontal dashed lines indicate the levels of no-obstacle case. 
}
\label{fig:Tgx_and_phi}
\end{figure}

To characterize the statistical property of $V_x$, here we introduce the granular temperature in horizontal direction, $T_{gx} \sim \langle \delta V_x^2 \rangle = \langle (V_x - \langle V_x \rangle )^2 \rangle$, where $\langle \cdot \rangle$ indicates the spatiotemporal average.
Since this granular temperature corresponds to the variance of PDF, it can be used to characterize the width of the distribution.
The measured $T_{gx}$ as a function of $L$ at $W=25$~mm is shown in Fig.~\ref{fig:Tgx_and_phi}(a).
As expected, $T_{gx}$ with the triangular-obstacle case is larger than the other obstacle cases in small $L$ regime.
The sudden increase of $T_{gx}$ at small $L$ comes from the detouring of particles around the obstacle.
The detouring followed by collisions below the obstacle results in the large horizontal velocity component.
Note that, however, $T_{gx}$ in the case of circular obstacle also significantly increases in the very small $L$ regime.
Since the clear and significant decrease of clogging occurence can be detected only for triangular obstacle at small $W$ and $L$ regime (Fig.~\ref{fig:phasediagram}), it is difficult to explain the anti-clogging mechanism solely by large $T_{gx}$.

\subsubsection{Packing fraction}
\label{subsubsec:Packing_fraction}
Another possible reason for reducing the clogging risk is the low packing fraction $\phi$. Here, $\phi$ is defined by the ratio of the particles area relative to the total area in the region {\sf{A}}. The measured $\phi$ as a function of $L$ at $W=25$~mm is shown in Fig.\,\ref{fig:Tgx_and_phi}(b). As expected, $\phi$ decreases as $L$ decreases.
Furthermore, the decreasing rate of $\phi$ is much more significant in the triangular-obstacle case than other two cases. This result is consistent with the fact that the triangular-obstacle case shows the clear reduction of clogging occurence in small $L$ and $W$ regime~(Fig.~\ref{fig:phasediagram}). Because the clogging reduction is clearly confirmed in $L<40$~mm (at $W=25$~mm) with the triangular obstacle~(Fig.~\ref{fig:phasediagram}(b)), we can consider that the sufficiently small $\phi (\leq 0.6)$ can safely prevent the clogging in two-dimensional granular flow through a narrow exit. This result is qualitatively consistent with~\cite{Roussel2007}. Obviously, the very small packing fraction is better to prevent clogging. In the very small $L$($\leq 25$~mm) regime, clogging by the arch formation between the obstacle and bottom wall~(Fig.~\ref{fig:clog}(b)) is observed. In this regime, $\phi$ in the region {\sf{A}} is no longer relevant to discuss te clogging. Rather, the packing fraction in the lateral side regions would be important in this regime. Namely, the important quantity is the packing fraction at the clogging (arch formation) region. In the case of inverted triangle, $\phi$ remains large $(\geq 0.65)$. Nevertheless, the inverted triangle can make smooth flow in some (seemingly random) regimes in the clogging diagram~(Fig.~\ref{fig:phasediagram}(c)). This point will be discussed later.

\begin{figure}
\begin{center}
\resizebox{0.45\textwidth}{!}{%
  \includegraphics{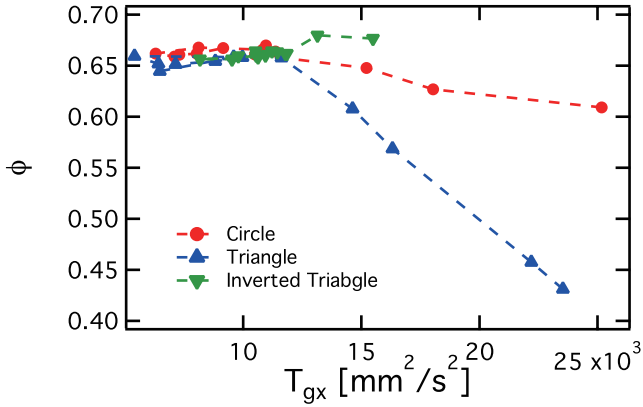} 
}
\end{center}
\caption{Correlation between packing fraction $\phi$ and horizontal granular temperature $T_{gx}$ at $W=25$~mm.}
\label{fig:phi_vs_Tgx}       
\end{figure}

Finally, we directly compare $\phi$ with $T_{gx}$ at $W=25$~mm. We can see the negative correlation between $\phi$ and $T_{gx}$ in large $T_{gx}$ regime~(Fig.~\ref{fig:phi_vs_Tgx}). Decreasing rate of $\phi$ becomes the maximum in the triangle case. This means that the triangular shape is the most efficient one to reduce $\phi$ by the identical $T_{gx}$. The flat bottom of the obstacle could play a crucial role to achieve this efficiency.
To investigate the details of this shape dependence, force visualization by photoelastic material~\cite{Tang2011,Vivanco:2012hc,Iikawa:2015ex,Iikawa:2016eg} and/or numerical simulation~\cite{Hidalgo:2013ko} might be helpful. In this study, we perform the numerical simulation as presented below.

\section{Numerical results and analyses}
\label{sec:numresults}

\begin{figure}
\begin{center}
\resizebox{0.45\textwidth}{!}{%
\includegraphics{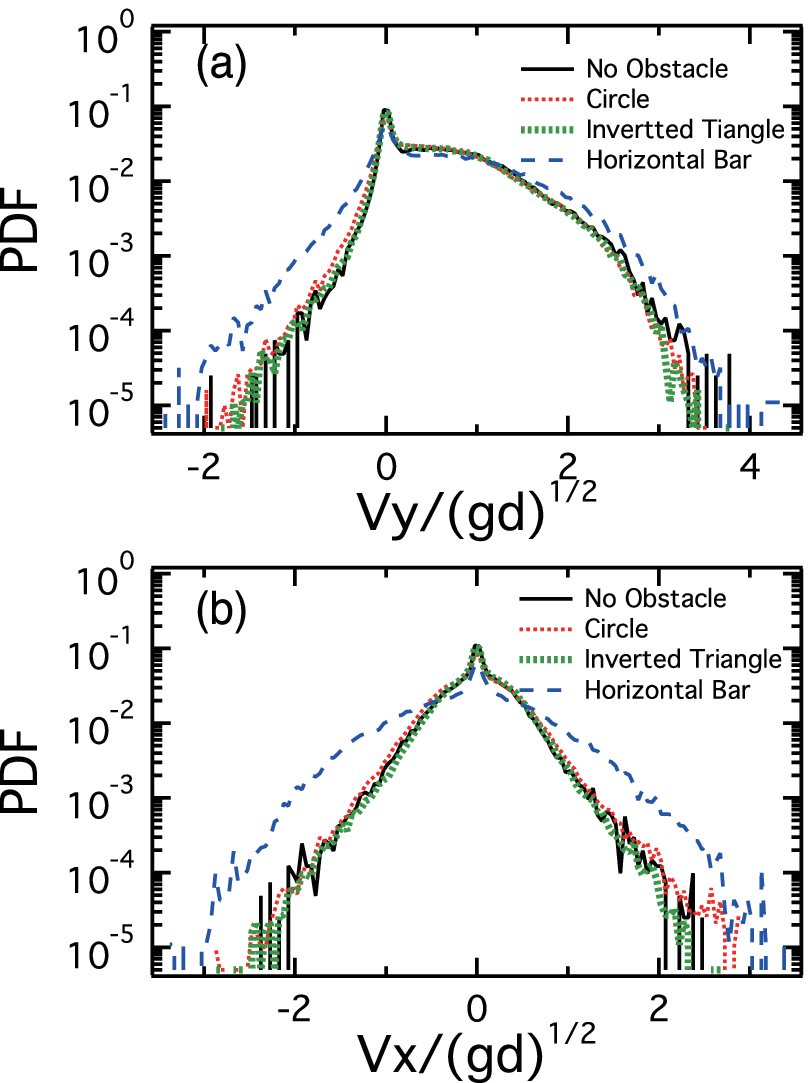}
}
\resizebox{0.45\textwidth}{!}{%
\includegraphics{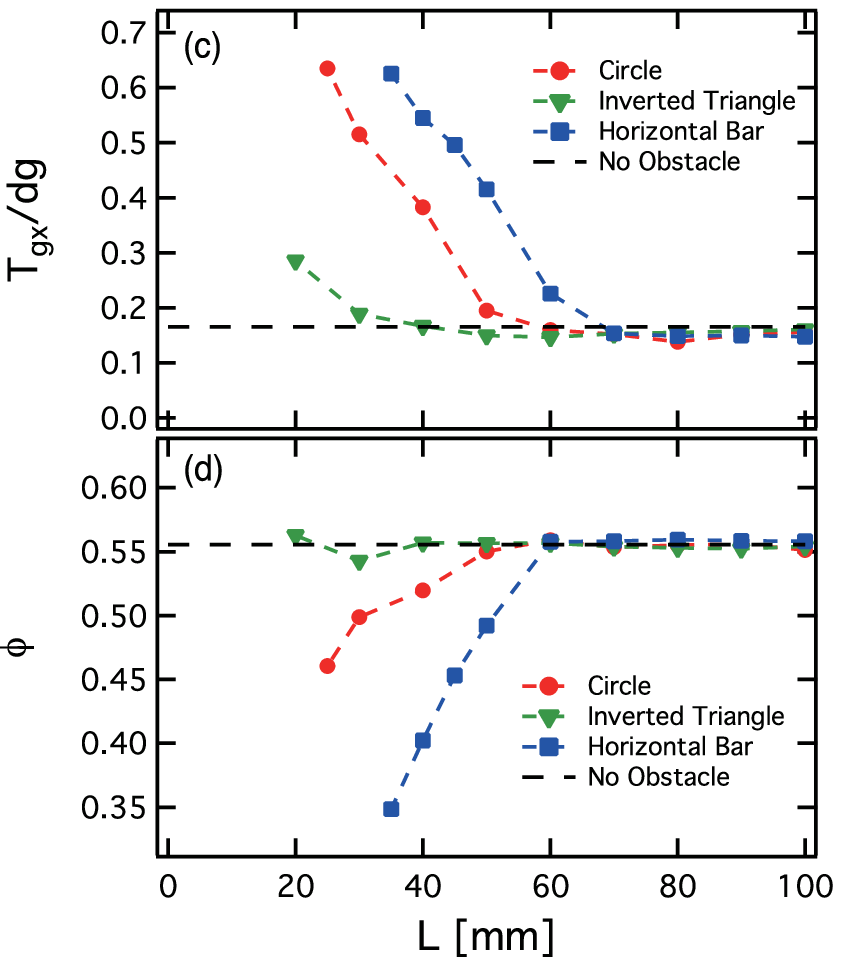}
} 
\end{center}
\caption{Numerically obtained data of (a) vertical velocity PDF, (b) horizontal velocity PDF, (c) horizontal granular temperature, and (d) packing fraction. For (a) and (b), $W=25$~mm and $L=50$~mm. For (c) and (d), $W=25$~mm.}
\label{fig:numresults} 
\end{figure}

Next, we compare the numerical result with experiment. In Fig.~\ref{fig:numresults}, numerically obtained PDF of $V_y$ and $V_x$, horizontal granular temperature $T_{gx}$, and packing fraction $\phi$ are shown. The parameter values of $W$ and $L$ used here are same as those used in the experiment. From the experimental observation, we guess that the particles detouring due to the obstacle and the horizontal bottom shape are the principal factors to make small $\phi$. Thus, in the numerical simulation, we employ a horizontal-bar obstacle instead of the triangle. We also examine the cases with circular and inverted-triangular obstacles. As can be seen in Fig.~\ref{fig:numresults}, almost all qualitative characteristics observed in the experiment are reproduced by the numerical simulation. Then, the horizontal bar in numerical simulation should correspond to the triangle in the experiment. Although the PDF shapes of $V_y$ and $V_x$ are slightly different from experimentally obtained ones, obstacle-shape dependence of the PDF shape and $L$ dependences of $T_{gx}$ and $\phi$ are basically captured by the numerical result. This result supports our speculation: particles detouring around the obstacle and horizontal bottom shape of the obstacle significantly reduce $\phi$ below the obstacle (at the exit region). And the reduced $\phi$ decreases the clogging risk (and flow rate). 

\section{Discussion}
\label{sec:discussion}
From the above observations, we suppose that the principal effect induced by the obstacle in granular discharge flow field is as follows.
By the presence of an obstacle at the vicinity of exit, particles have to detour to approach the exit.
Thereby, particles-supply rate to the exit region is decreased by the detouring effect. This effect directly reduces local packing fraction at the exit region. The most significant detouring can be made by triangular obstacle (or horizontal bar). 
In addition, at the central part beneath the obstacle, strong collisions occur among the particles coming from the left and right sides.
Then, the random motion of particles, which corresponds to the ``temperature'' in this system (granular temperature), is increased.
Finally, the packing fraction right above the exit is further reduced by the excluded volume effect of active (high temperature) particles.
As a consequence, the resultant low packing fraction prevents the arch formation at the exit.
Thus, these obstacle effects can avoid the clogging occurrence.
In this sense, the obstacle is useful to prevent the clogging by making small $\phi$.

This simple scenario to prevent the clogging cannot explain all the situations. For instance, when the inverted-triangular obstacle is used, the clogging can be observed rather randomly at $W=25$~mm~(Fig.~\ref{fig:phasediagram}(c)) in spite of large $\phi$. This result implies that the clogging prevention can be achieved even in large $\phi$ regime. We speculate that the crystal-like ordered structure made at the exit region could be a key to keep large $\phi$ by the inverted triangle~(Fig.~\ref{fig:flowfield}(g)). Although such an ordered structure yields the large $\phi$, its structure could flow systematically forming a cluster without developing an arch, as long as it does not create the obstacle-wall arch like Fig.~\ref{fig:simuexample}(b). Then, the competition between high density and clustering might result in a very stochastic (not systematic) clogging as observed in the inverted-triangular case at $W=25$~mm. Moreover, the pressure decrease above the obstacle will also be effective to prevent the clogging~\cite{Zuriguel2011}. Namely, there might be plural mechanisms to prevent the clogging by the obstacle. Detail classification and characterization of various obstacle effects are still important future problems.

There are some interesting issues we have not discussed in this paper. For example, spatiotemporal analyses of alternate flow typically observed in circular- and triangular-obstacle cases have not been performed. Actually, the experimental conditions to reproduce the alternate flow could not be precisely determined from our experimental result. It occurred rather in a random fashion. Perhaps, its classification might be more complex than making the clogging diagram. And the preliminary-measured switching period of the alternate flow also seemed to be neither universal nor systematic. Its characterization is one of the challenging issues opened to future study. In addition, the forms of PDF (Fig.~\ref{fig:PDF}) have not been discussed in detail. Only $T_{gx}$ has been used in this study. Its detail characterization is also an interesting future problem.

\section{Conclusion}
\label{sec:Conclusion}
We conducted simple experiment and numerical simulation of two-dimensional granular flow driven by gravity under the influence of an obstacle.
From the images of the granular exit flow acquired by a high-speed camera, we tracked the motion of individual particles at the exit region by means of PTV method.
By the triangular obstacle, the clogging occurence can be reduced at a certain distance range from the exit.
The principal reason for this clogging prevention is the low packing fraction at the vicinity of exit.
The free space among the particles seems to be effectively created when the triangular obstacle approaches the exit.
At the same time, the triangular obstacle also results in the large granular temperature in horizontal direction.
The detouring by the triangle can decrease the supply of the particles to the exit region, and also activate the particles by collisions to have large velocity component in a horizontal direction.
These behaviors were reproduced by the numerical simulation as well. And, the numerical simulation also revealed that the horizontal bar is enough to make the above-mentioned situations: small $\phi$ below the obstacle (above the exit region).

\section*{Acknowledgement}
We would like to acknowledge S.~Watanabe, H.~Kumagai, S.~Sirono, and T.~Morota for helpful discussion. This work has been supported by JSPS KAKENHI No.~15H03707. K. Anki Reddy would like to thank IITG start-up research grant.

\bibliography{clog}

\end{document}